\documentclass[%
aip,
amsmath,amssymb,
reprint,%
]{revtex4-1}
\usepackage{amsmath}
\usepackage{graphicx}
\usepackage{bm}
\usepackage{hyperref}
\usepackage{multirow}

\newcommand{\tref}[1]{Table~\ref{#1}}

\begin{document}
	\title{Search for CP-violating nuclear magnetic quadrupole moment using the LuOH$^+$ cation}

	\begin{abstract}
		The time-reversal and spatial parity violating interaction of the nuclear magnetic quadrupole moment (MQM) of the $^{175}$Lu and $^{176}$Lu nuclei with electrons in the molecular cation LuOH$^+$ is studied. The resulting effect is expressed in terms of fundamental parameters, such as quantum chromodynamics angle $\bar{\theta}$, quark electric dipole moment (EDM) and chromo-EDM. For, this we have estimated magnetic quadrupole moments of $^{175}$Lu and $^{176}$Lu nuclei and calculated the molecular constant that characterises the interaction of the MQM with electrons in the considered molecules. Additionally, we predict the hyperfine structure constants for the ground electronic state of LuOH$^+$. In the molecular calculations, both the correlation and relativistic effects including the Gaunt interaction have been considered. According to the calculated expressions in terms of the fundamental constants, we conclude that LuOH$^+$ can be a promising system to measure the nuclear MQM.
	\end{abstract}

	\author{D.E. Maison}
	\affiliation{Petersburg Nuclear Physics Institute named by B.P.\ Konstantinov of National Research Center ``Kurchatov Institute'' (NRC ``Kurchatov Institute'' - PNPI), 1 Orlova roscha mcr., Gatchina, 188300 Leningrad region, Russia}
	\affiliation{Saint Petersburg State University, 7/9 Universitetskaya nab., St. Petersburg, 199034 Russia}
	\homepage{http://www.qchem.pnpi.spb.ru    }
	\email{daniel.majson@gmail.com, maison\_de@pnpi.nrcki.ru}
	
	\author{L.V.\ Skripnikov}
	\affiliation{Petersburg Nuclear Physics Institute named by B.P.\ Konstantinov of National Research Center ``Kurchatov Institute'' (NRC ``Kurchatov Institute'' - PNPI), 1 Orlova roscha mcr., Gatchina, 188300 Leningrad region, Russia}
	\affiliation{Saint Petersburg State University, 7/9 Universitetskaya nab., St. Petersburg, 199034 Russia}

	\author{V.V.\ Flambaum}
	\affiliation{School of Physics, The University of New South Wales, Sydney NSW 2052, Australia}
	\affiliation{Johannes Gutenberg-Universit\"at Mainz, 55099 Mainz, Germany}
	
	\author{M. Grau}
	\affiliation{Institute for Quantum Electronics, ETH Z{\"u}rich, Otto-Stern-Weg 1, 8093 Z{\"u}rich, Switzerland}
	
	\maketitle
	
	\section{Introduction}
	
	Study of the time-reversal (T) and spatial parity (P) violating forces inside the nucleus is of the key importance to test extensions of the Standard Model of elementary particles.
	According to the CPT theorem, all the observable phenomena must be symmetric with respect to the combined CPT-transformation, where C is the charge conjugation. 
	This statement was verified in the recent experiments with an uncertainty lower than 1~ppb~\cite{Safronova:18}.
	%dm_resp2
	Therefore,
	%end dm_resp
	T-violation implies necessarily also CP-violation. As it was shown by Sakharov \cite{Sakharov1967}, CP-invariance violation is one of the three necessary conditions to explain the baryon asymmetry of the Universe, also known as matter-antimatter asymmetry. Investigation of CP-violation in nature is closely related to the explanation of the predominance of baryon matter, which is called ``one of the great mysteries in physics'' \cite{canetti2012matter} and can not be explained within the Standard Model.
	
	The nonzero value of the electric dipole moment (EDM) of a system (e.g., a free particle) with nonzero spin $\mathbf{S}$ would imply violation of both T- and P-symmetries. This is a consequence of the Wigner-Eckart theorem, where the mean value of the EDM vector should be proportional to the mean value of the spin vector. However, the dipole moment vector is P-odd and T-even while the spin vector is P-even and T-odd.
	The EDM of a system can serve as an indicator of the T,P-symmetry violation and 
	can be induced by different T,P-odd interactions inside this system.
	
	In the second half of the twentieth century, it was realized that atoms and molecules with heavy atoms are
	very promising systems to search for the T,P-invariance violation effects~\cite{Safronova:18,KL95}.
	Indeed, the best constraints on the electron electric dipole moment (eEDM) have been established in the molecular experiment using the neutral $^{232}$ThO molecule~\cite{ACME:18}. A very sensitive experiment to search for the electron EDM has also been performed using the trapped $^{180}$HfF$^+$ molecular cations~\cite{Cornell:2017}.
	
	The present study is devoted to another source of CP-violation, the nuclear magnetic quadrupole moment (MQM). It is induced by CP-violating internal nuclear interactions and unlike the eEDM therefore can serve as an indicator of the CP-invariance violation in the hadron sector. 
	As it is shown below, nuclear MQM is determined by the CP-violating parameters 
	of the Standard Model, such as
	(i)
	vacuum angle $\bar{\theta}$, the free parameter of the theory, which can have an arbitrary value from the interval $\left(-\pi; \pi\right)$ but due to unknown reasons is extremely close to zero, and 
	(ii)
	$u$- and $d$-quarks chromo-EDM, which are analogs of electric dipole moment for strong interactions.
	Therefore, MQM measurement can be used to determine these parameters, which are currently unknown. 
	It is important to note that in deformed nuclei like $^{175}$Lu and $^{176}$Lu considered in present study,
	MQM has collective nature and is enhanced by an order of magnitude \cite{flambaum1994spin}. 
	According to the Standard Model and popular 
	quantum chromodynamics (QCD) $\theta$-term model the MQM contribution is several orders of magnitude greater than the electron EDM contribution \cite{PospelovRitzreview}.
	Within the Standard Model MQM arises due to the CP-violating phase in the quark mass matrix \cite{Sushkov:84}.
	However, the main goal of MQM measurement is the  search for physics beyond the Standard Model including the test of unification theories.
	For example, axion dark matter produces an oscillating $\theta$-term~\cite{Graham2011}, therefore, a MQM experiment may be used to search for dark matter~\cite{Stadnik2014}.
	
	The symmetry violating nature of MQM can be seen~\cite{Ginges:04} by isolating the MQM term in the multipole expansion of the vector potential $\mathbf{A}$ produced by a steady current inside a nucleus in the Coulomb gauge $(\nabla, \mathbf{A})=0$. This procedure leads to the following expression for MQM tensor components:
	\begin{equation}
	\label{Mim}
	M_{i,m}=-\int (r_i \epsilon_{mnp}+r_m \epsilon_{inp})j_n r_p d^3r,    
	\end{equation}
	where $j$ is the vector current density. At the same time MQM should be proportional to the only irreducible rank-2 tensor $T_{i,k}$ constructed from the total angular momentum $\mathbf{I}$ of the system, $T_{i,k}=I_i I_k+ I_k I_i -\frac{2}{3}I(I+1)\delta_{ik}$. 
	By comparing these expressions one concludes that MQM violates both T and P symmetries.
	If we consider a paramagnetic atom or a molecule with nonzero  electronic angular momentum, the electrons of the system produce a magnetic field interacting with MQM. 
	This interaction mixes states of opposite parity and generates atomic or molecular EDM.
	
	The basic principle of the MQM experiment is similar to the principle of the existing molecular EDM and Schiff moment experiments. The molecule is polarized in some lab frame electric field, and a Ramsey measurement is performed between two spin states in an applied magnetic field. The hypothetical T,P-violating shifts appear on top of the measured Zeeman shift. The separation from the Zeeman shift is achieved by changing sign of electric  field $\textbf{E}$ relative to magnetic field $\textbf{B}$, i.e. change of sign of the product  $(\textbf{E} \cdot  \textbf{B})$. The T,P-violating shift is proportional to $\textbf{E}$ and changes sign.  
	The exact way this Ramsey measurement is done can vary slightly from apparatus to apparatus. For example, in the ThO experiment the spin state is read out using polarized laser induced fluorescence, and in the HfF$^+$ experiment this is done using resonant multiphoton dissociation. In a single molecular ion LuOH$^+$ experiment it might be possible to perform the Ramsey measurement using quantum logic spectroscopy with a co-trapped Lu$^+$ atomic ion.
	
	Recently, it was noted~\cite{isaev2016polyatomic, kozyryev2017precision} that linear triatomic molecules with heavy atoms such as YbOH have certain advantages to search for T,P-violating effects over corresponding isoelectronic diatomic molecules. In such molecules, there is a small energy gap between levels of opposite parity due to the $l$-doubling effect~\cite{kozyryev2017precision}. This allows one to fully polarize them using a relatively weak electric field. Additionally, the $l$-doublet can be used to suppress systematic errors arising from magnetic fields, regardless of the electronic state. Another feature of linear triatomics is their ability to be cooled by the laser-cooling technique. The ytterbium monohydroxide molecule has been studied by several theoretical groups \cite{Maison:2019b, denis2019enhancement,gaul2018ab,prasannaa2019enhanced,Maison:2020b}. It can be cooled to temperatures lower than 1 mK \cite{augenbraun2020laser}. This allows one to increase the coherent time significantly and 
	decrease uncertainty of the experiment, which is inversely proportional to this time. 
	
	In the present paper, we study other triatomic systems ---  the $^{175}$LuOH$^+$ and $^{176}$LuOH$^+$ molecular cations. 
	They have the same aforementioned advantages of triatomic molecules, and additional benefits inherent to ions: long storage times are possible in an ion trap, and a co-trapped atomic ion can be used to sympathetically cool the molecule, as well as to initialize and read out its state using quantum logic spectroscopy~\cite{Wolf2016}.
	The  $^{175}$Lu nucleus has spin $I=7/2$, $^{176}$Lu has spin $I=7$ and they can therefore have the CP-violating nuclear MQM. Natural abundance of $^{175}$Lu is 97.4 \% and abundance of $^{176}$Lu is 2.6 \%. However, as we see below, MQM  of $^{176}$Lu is 1.4 times bigger than $^{175}$Lu. Experimentally, $^{175}$Lu and $^{176}$Lu are appealing species as the atomic ion Lu$^+$ can be directly laser cooled and stored under ultra-high vacuum in an ion trap~\cite{Kaewuam2018}. The molecular ion LuOH$^+$ could then be created by reacting the cold Lu$^+$ with, e.g., water or methanol. Remaining Lu$^+$ ions can be used to sympathetically cool the LuOH$^+$ molecules. The LuOH$^+$ molecule has a simple electronic structure, with a single valence electron and a $^2\Sigma_{1/2}$ ground state, which facilitates state preparation via optical pumping schemes, and state detection of the molecule can be achieved by performing resonant dissociation followed by observing Lu$^+$ fluorescence, or via quantum logic spectroscopy using a co-trapped Lu$^+$ as a logic ion.

	The T,P-violating energy shift that can be measured in LuOH$^+$ can be induced by the MQM of the Lu nucleus. It depends on the MQM value itself and the molecular interaction parameter $W_M$ (see below).
	MQM can be further expressed in terms of fundamental interaction parameters.
	We provide both nuclear and many-body electronic structure calculations to express the expected 
	T,P-violating
	effect in terms of fundamental interaction parameters.
	According to our findings the characteristic molecular enhancement parameter $W_M$ for LuOH$^+$ cation appears to be 16\% higher than in the isoelectronic neutral molecule YbOH. 
	More importantly, MQM of $^{175}$Lu is $\sim$ 30\%  bigger and MQM of  $^{176}$Lu is $\sim$ 70\% bigger than MQM of $^{173}$Yb (see below). Altogether, the size of a T,P-violating energy shift in $_{71}^{175}$LuOH$^+$ and $_{71}^{176}$LuOH$^+$ may be a factor of 2 higher than in  $_{70}^{173}$YbOH.
	
	In addition we calculate magnetic dipole hyperfine structure (HFS) constants, the measurement of which may be used as some test of the accuracy of the electronic structure calculation (see also a discussion on such tests in Ref.~\onlinecite{Skripnikov:2020e}). 
	%end ls_resp

	\section{Nuclear magnetic quadrupole moment}
	
	We perform calculation of MQM using the technique used in Ref. \onlinecite{lackenby2018time}. The $^{175}$Lu nucleus is deformed, therefore, we use the deformed oscillator Nilsson model for proton and neutron orbitals. Formula for the contribution of a Nilsson orbital to the nuclear MQM has been derived in Ref. \onlinecite{lackenby2018time}: $M^{p,n}= 4 \Sigma \Lambda M_0^{p,n}$, where $\Sigma$ and $\Lambda$ are projections of the nucleon spin and orbital angular momentum on the nuclear axis,   $M^p_0$  and  $M^n_0$ are the single-particle matrix elements for protons and neutrons  which depend on the form of the T,P-odd interaction (see below).   
	Summation over nucleons  gives the following result for the  $^{175}$Lu collective MQM:
	\begin{equation}
	\label{M}
	M= 15 M^p_0 + 32 M^n_0,
	\end{equation}
	MQM for $^{176}$Lu is bigger:
	\begin{equation}
	\label{M176}
	M= 21 M^p_0 + 40 M^n_0,
	\end{equation}
	Note that the ratio of MQM for $^{176}$Lu and for $^{175}$Lu is approximately equal to the ratio 1.4 of the measured electric quadrupole moments of these nuclei presented in Tables \cite{stone2016table}. This is not accidental. Both MQM and electric quadrupole have collective nature. In the rotating frame (frozen body frame) one extra neutron in $^{176}$Lu does not play an important role since the  number of nucleons contributing to the collective quadrupoles is large, $\sim A^{2/3}$. However, transition to the laboratory frame adds the factor $\frac{I(2I-1)}{(I+1)(2I+3)}$, where the nuclear spin $I=7$ for $^{176}$Lu and $I=7/2$ for $^{175}$Lu. The ratio of the angular factors for $^{176}$Lu and  $^{175}$Lu is 1.4.  Therefore, in the following section we only present the results for $^{175}$Lu. Results for  $^{176}$Lu may be obtained by multiplying all numbers for $^{175}$Lu by 1.4.

	For comparison we also calculated $^{173}$Yb collective MQM:
	\begin{equation}
	\label{MYb}
	M= 14 M^p_0 + 23 M^n_0,
	\end{equation}
	The coefficient before the neutron contribution (equal to 23) is now slightly different from the value in Ref.~\onlinecite{lackenby2018time} (formerly 26) due to a small contribution of deep neutron orbitals accounted for in the present work. 
	%The difference is within the accuracy of the nuclear calculations.
	
	We start from a contact T,P-odd nuclear potential 
	\begin{equation}
	\label{VTP}
	V^{TP}_{p,n}= \eta_{p,n}  \frac{G} {2^{3/2} m_p} (\sigma \cdot \nabla \rho),
	\end{equation}
	acting on the valence nucleon. 
	Here  $\rho$ is the total nucleon number density, $G$ is the Fermi constant, $m_p$ is the proton mass, $\eta_{p}$  and $\eta_{n}$ are the dimensionless strength constants for protons and neutrons (note that  constants  $\eta_{p}$  and $\eta_{n}$ in the nucleon-nucleus potential contain weighted sum over different nucleon-nucleon interaction constants since the nuclear density $\rho$ contains sum over nucleons).  Using Eq. (\ref{M}) and values of the single-particle matrix elements $M^p_0=-0.76 \eta_p \cdot 10^{-34} e \cdot  \text{cm}^2 + 2.1 d_p \cdot 10^{-14}  \text{cm}$  and  $M^n_0= 0.80 \eta_n \cdot 10^{-34} e \cdot  \text{cm}^2 + 2.1 d_n \cdot 10^{-14}  \text{cm}$  from  Refs. \onlinecite{FDK14,lackenby2018time} we obtain for $^{175}$Lu MQM:
	\begin{multline}
	\label{Meta}
	M= (2.6 \eta_n - 1.1 \eta_p) \cdot 10^{-33} e \cdot  \text{cm}^2 +\\
	+(0.67 d_n + 0.31 d_p) \cdot 10^{-12}  \text{cm},
	%\frac{\hbar}{m_p c} ,
	\end{multline}
	where $d_n$ and $d_p$ are neutron and proton electric dipole moments.
	%$\lambda_p=\hbar /m_pc=2.10 \cdot 10^{-14}$ cm.
	The T,P- odd nuclear potential Eq. (\ref{VTP}) is dominated  by the neutral $\pi_0$ exchange between the nucleons and the strength constants $\eta$ may be expressed in terms  of $\pi NN$ couplings (see details in  Ref. \onlinecite{FDK14}):
	\begin{align}
	\eta_{n} = -\eta_{p} \approx 5\cdot 10^{6}g\left(\bar{g}_1 + 0.4 \bar{g}_2\ - 0.2\bar{g}_0\right) ,
	\end{align}
	where $g$ is the strong $\pi NN$ coupling constant  and  $\bar{g}_0\,, \bar{g}_1\,, \bar{g}_2$ are  three T,P-odd $\pi NN$ coupling constants, corresponding to the different isotopic  channels.  Substitution of these $\eta_{n,p}$ into Eq. (\ref{Meta}) gives:
	\begin{eqnarray}
	\label{Mg}
	\nonumber
	M= g \left(1.8 \bar{g}_1 + 0.73 \bar{g}_2\ - 0.37\bar{g}_0\right) \cdot 10^{-26} e \cdot  \text{cm}^2 \\
	+ (0.67 d_n + 0.31 d_p) \cdot 10^{-12}  \text{cm},
	%\frac{\hbar}{m_p c} ,
	\end{eqnarray}
	Constants of the T,P-odd $\pi NN$ interaction  $\bar{g}$ and nucleon EDMs may be expressed in terms of more fundamental T,P- violating parameter,  QCD constant  $\bar{\theta}$, or EDM  $d$ and chromo-EDM $\tilde{d}$ of $u$ and $d$ quarks
	%ls1
	~\cite{Bsaisou:15,Dekens:14,theta}.
	Definitions of these parameters are given by Eqs. (1), (2) and (3) of Ref.~\onlinecite{yamanaka2017}, respectively.
	
	$$g\bar{g}_0(\bar{\theta}) =  -0.21 \bar{\theta}$$ %typo?
	%end ls
	$$g\bar{g}_1(\bar{\theta}) =  0.046 \bar{\theta}$$
	$$d_n=-d_p=1.2 \cdot 10^{-16} \bar{\theta} \cdot  e \cdot \text{cm}$$
	$$g\bar{g}_0(\tilde{d}_u, \tilde{d}_d)= 0.8\cdot 10^{15} \left(\tilde{d}_u +\tilde{d}_{d}\right) \ \text{cm}^{-1} $$
	$$g\bar{g}_1(\tilde{d}_u, \tilde{d}_d)= 4\cdot 10^{15} \left(\tilde{d}_u  - \tilde{d}_{d}\right) \ \text{cm}^{-1} $$
	$$d_{p}(d_u, d_d, \tilde{d}_u, \tilde{d}_d) = 1.1e\left(\tilde{d}_u + 0.5\tilde{d}_{d}\right) + 0.8 d_u - 0.2d_d $$
	$$d_{n}(d_u, d_d, \tilde{d}_u, \tilde{d}_d) = 1.1e\left(\tilde{d}_d + 0.5\tilde{d}_{u}\right) - 0.8 d_d + 0.2d_u $$
	
	The substitutions to Eq.  (\ref{Mg}) give the following results for  $^{175}$Lu MQM:
	\begin{align}  \label{Md}
	M(\bar{\theta}) \approx 1.6 \cdot 10^{-27} \, \bar{\theta} e \cdot  \text{cm}^2 \,, \\
	\label{MdD}
	M(\tilde{d}) \approx   0.7  \cdot 10^{-10 }  \left(\tilde{d}_u  - \tilde{d}_{d}\right) e \cdot  \text{cm}   
	\end{align}
	The  $^{176}$Lu MQM is 1.4 bigger, so we have coefficients $2.2 \cdot 10^{-10 }$ in Eq. (\ref{Md}) and $1.0 \cdot 10^{-10 }$ in Eq. (\ref{MdD}). For comparison, 
	according to Ref. \onlinecite{Maison:2019b} for $^{173}$Yb coefficients are $1 \cdot 10^{-10 }$ and $0.6 \cdot 10^{-10 }$ correspondingly.

	\section{Electronic structure}
	
	\subsection{Geometry optimization}
	Nuclear configuration of triatomic molecule has three degrees of freedom: two interatomic bond lengths and one bond angle. According to our calculations
	performed in this paper, the LuOH$^+$ cation has a linear geometry in the ground open-shell 
	electronic state $^2\Sigma_{1/2}$. Taking this into account, one can determine equilibrium geometry parameters as a minimum point of the energy function $E = E(R(\textrm{Lu--O}),R(\textrm{O--H}))$.
	In the present paper we have performed such minimization by the numerical technique. 
	Solution of the electronic problem for each set of 
	parameters $R(\textrm{Lu--O})$ and $R(\textrm{O--H})$
	arising in the optimization process has been performed within the relativistic 4-component coupled cluster approach with single, double and perturbative triple cluster amplitudes CCSD(T) \cite{Bartlett1991,Crawford:00,Bartlett:2007}. The inner electrons of Lu with lowest orbital energies (1s$^2$2s$^2$2p$^6$) were excluded from this correlation calculation, as well as virtual orbitals with  energies greater than 600 Hartree. For comparison, the orbital energy of 3s electrons of the Lu atom, which is the lowest active shell, is $-94$ Hartree. Dependence of electronic properties on energy cutoff was studied extensively in Refs.~\onlinecite{Skripnikov:17a,Skripnikov:15a}. In the calculations the  Gaussian-type basis set has been employed. The uncontracted Dyall's CV3Z basis set for Lu atom \cite{gomes2010relativistic} and the aug-cc-PVTZ-DK basis sets \cite{Dunning:89,Kendall:92,de2001parallel} for the oxygen and hydrogen atoms were used. Electronic structure calculations reported in this paper were performed using the local version of the {\sc dirac15} \cite{DIRAC15} and {\sc mrcc} codes~\cite{MRCC2020,Kallay:1,Kallay:2}. 
	The determined equilibrium geometry parameters are: $R(\textrm{Lu--O}) =1.873(20) \textrm{\r{A}}$ and $R(\textrm{O--H}) = 0.958(20) \textrm{\r{A}}$.

	\subsection{$W_M$ calculation}
	The T,P-violating interaction of the nuclear MQM with electrons is described by the following Hamiltonian:
	\begin{equation} \label{HmqmOperator}
	H_{MQM} 
	=
	-\frac{M}{2I\left(2I-1\right)} T_{i,k} \cdot \frac{3}{2} \frac{\left[\boldsymbol{\alpha}\times \mathbf{r}\right]_i r_k}{r^5},
	\end{equation}
	where $T_{i,k}=I_i I_k+ I_k I_i -\frac{2}{3}I(I+1)\delta_{ik}$, $I$ is the nuclear spin of 
	heavy nucleus
	$M$ is 
	its
	magnetic quadrupole moment,
	$\boldsymbol{\alpha}$ are Dirac matrices and $\mathbf{r}$ is the electron radius-vector with respect to the heavy atom nucleus under consideration. 
	Nucleus-dependent
	T,P-violating effects may also be produced by the Schiff moment and electric octupole moment \cite{Sushkov:84,Flambaum:97}. We do not consider them here since for Lu and Yb nuclei (where MQM has collective enhancement) the Schiff moment and electric octupole moment effects are two orders of magnitude smaller than that of MQM. 
	
	The electronic part of the Hamiltonian (\ref{HmqmOperator}) is characterized by the molecular constant $W_M$ \cite{Sushkov:84,Dmitriev:92,Skripnikov:14a}:
	\begin{equation}
	\label{WmME}
	W_M
	=
	\frac{3}{2 \Omega}
	\langle \Psi | \sum\limits_i 
	\left(
	\frac{\boldsymbol{\alpha}_i\times\mathbf{r}_i}
	{r_i^5}
	\right)_\zeta r_\zeta | \Psi \rangle,
	\end{equation}
	where $\Psi$ is the electronic wavefunction, index $i$ runs over all the electrons, index $\zeta$ means projection on the molecular axis and $\Omega$ is the projection of the total electronic angular momentum $\mathbf{J}^e$ on the molecular axis. The ground electronic state of the LuOH$^+$ cation has $\Omega = 1/2$. $W_M$ constant can not be measured but is required for interpretation of the experimental data in terms of the nuclear MQM.  In order to test the accuracy of the obtained $W_M$ value we have performed calculation of the ground electronic state hyperfine structure constants, which can be measured directly (see below).
	
	\subsection{$A_{||}$ and $A_{\perp}$ calculation}
	Magnetic dipole hyperfine structure of the $^2 \Sigma_{1/2}$ state is described by the following constants:
	\begin{equation} \label{HyperfineSplitting}
	A_{||} = \frac{\mu_{\rm Lu}}{\Omega I_{\rm Lu}}
	\langle \Psi_{^2 \Sigma_{+\frac{1}{2}}} | 
	\sum\limits_i  
	\left(\frac{\mathbf{r}_i \times \boldsymbol{\alpha}_i}{r_i^3}\right)_\zeta 
	| \Psi_{^2 \Sigma_{+\frac{1}{2}}} \rangle,
	\end{equation}
	
	\begin{equation} \label{A_perp}
	A_{\perp} = \frac{\mu_{\rm Lu}}{I_{\rm Lu}}
	\langle \Psi_{^2 \Sigma_{+\frac{1}{2}}} | 
	\sum\limits_i  
	\left(
	\frac{\mathbf{r}_i \times \boldsymbol{\alpha}_i}
	{r_i^3}
	\right)_+
	| \Psi_{^2 \Sigma_{-\frac{1}{2}}} \rangle.
	\end{equation}
	
	Here $\mu_{\rm Lu}$ is the magnetic moment of the Lu nucleus, $I_{\rm Lu}$ is its spin, index ``+'' denotes the following linear combination: $a_+ = a_x + i a_y$ (and similarly for other vectors), where $xy$-plane is perpendicular to the molecular axis. Below we will also use the nuclear $g$ factor of the Lu nucleus, $g_{\rm Lu}=\mu_{\rm Lu}/I_{\rm Lu}$.
	As one can see from Eqs. (\ref{WmME}) -- (\ref{A_perp}), the values of $W_M$, $A_{||}$ and $A_{\perp}$ constants are  mainly determined by the behavior of the 
	wavefunction of the unpaired electrons in the vicinity of the heavy atom nucleus.
	Considered characteristics are examples of so-called atoms-in-compounds (AIC) properties~\cite{Titov:14a,Skripnikov:15b}. 
	Note, that correlation contributions from different electronic shells to AIC characteristics usually have a similar structure~\cite{Skripnikov:15a,Skripnikov:2020e}. Therefore, it is possible to obtain an indirect estimation of the uncertainty of the $W_M$ value by calculating the hyperfine structure constant~\cite{Quiney:98,Titov:06amin,Skripnikov:15b,Skripnikov:15a,Sunaga:16,Fleig:17,Borschevsky:2020,Skripnikov:2020e}. Thus, 
	in the point nuclear magnetic dipole model $A_{||}/g_{\rm Lu}$ and $A_{\perp}/g_{\rm Lu}$ parameters are independent of the isotope considered. The hyperfine splitting constants may be obtained by their scaling with the factor $g_{^{175}\textrm{Lu}} = 0.6379$ and $g_{^{176}\textrm{Lu}} = 0.4527$ \cite{Mills:93}. The $A_{||}$ and $A_{\perp}$ parameters can be used to estimate indirectly the $W_M$ uncertainty if the value of hyperfine splitting is measured~\cite{Kozlov:97c,Skripnikov:15b,Titov:06amin}.
	
	To compute matrix elements (\ref{WmME})--(\ref{A_perp}) the code developed in Refs.~\onlinecite{Skripnikov:17b,Skripnikov:16b} was used.

	\section{Results and discussion}
	In order to check the convergence of the $W_M$ and hyperfine constants values with respect to the basis set size we have performed calculations using five basis sets. 
	%dm
	The values of the considered parameters calculated using these basis sets are given in \tref{TResult1} in order of the increase of the basis set quality.
	%end dm
	%ls
	Note that in all cases the Dyall's all-electron basis sets~\cite{gomes2010relativistic} (AE2Z, AE3Z and AE4Z) were taken in the uncontracted form. Basis sets of the aug-cc-pV(D,T)Z family~\cite{Dunning:89,Kendall:92,de2001parallel} were taken in the contracted form.
	Inner core electrons $1s^22s^22p^6$ of Lu were excluded from these calculations. Energy cutoff of virtual orbitals was set to 1000 Hartree in these calculations. 
	As one can see, except the smallest basis set case, the $W_M$ and $A_{||}$ values are weakly dependent on the basis size.
	According
	to \tref{TResult1} increase of the basis set size on Lu from the AE3Z to the AE4Z one with the fixed aug-cc-pVTZ basis sets on light atoms changes the $W_M$ and $A_{||}$ values by about 0.2\%.
	The variation of O\&H basis sets with the fixed lutetium basis set also affects these parameters less than 0.5\%.
	Therefore, the uncertainty of the final results due to the basis set incompleteness can be estimated to be less than 1\%.
	
	\begin{table}
		\caption{
			Dependence of the calculated values of the $W_M$,  $A_{||}/g_{\rm Lu}$ and $A_{\perp}/g_{\rm Lu}$ parameters for the ground electronic state of  $^{175}$LuOH$^+$ on different basis sets within the relativistic CCSD(T) approach; $1s^22s^22p^6$ electrons of Lu were excluded from correlation treatment.}
		\label{TResult1}
		\begin{tabular}{lllll}
			\hline
			\hline
			Basis set $ \ \  $  & basis set  $ \ \   $  & $W_M$,        &   $A_{||}/g_{\textrm{Lu}}$,  & $A_{\perp}/g_{\textrm{Lu}}$\\
			on Lu \cite{gomes2010relativistic}     $  \ \ $  & on O and H \cite{Dunning:89,Kendall:92,de2001parallel}$  \ \ $  & 10$^{33} \frac{\textrm{Hz}}{e\cdot \textrm{cm}^2}$ &    MHz & MHz\\
			\hline
			AE2Z & aug-cc-pVDZ-DK & $-$1.224 & 12356 & 11942\\
			AE3Z & aug-cc-pVDZ-DK & $-$1.248 & 12558 & 12133\\
			AE3Z & aug-cc-pVTZ-DK & $-$1.251 & 12579 & 12153\\
			AE4Z & aug-cc-pVDZ-DK & $-$1.244 & 12560 & \\
			AE4Z & aug-cc-pVTZ-DK & $-$1.249 & 12582 & \\
			\hline
			\hline
		\end{tabular}
	\end{table}

	The final calculated values of the $W_M$, $A_{||}/g_{\rm Lu}$ and $A_{\perp}/g_{\rm Lu}$ constants are given in Table~\ref{TResult2}. The main calculation has been performed using the basis set that corresponds to the uncontracted AE3Z basis set on Lu \cite{gomes2010relativistic} and aug-cc-pVTZ-DK \cite{Dunning:89,Kendall:92,de2001parallel} on light atoms. All electrons were correlated in this calculation and the energy cut-off for virtual orbitals was set to 11000 Hartree~\cite{Skripnikov:17a,Skripnikov:15a}. We have also applied the basis set correction calculated as the difference between results obtained using the AE4Z(Lu)\&aug-cc-pVTZ-DK(O,H) and AE3Z(Lu)\&aug-cc-pVTZ-DK(O,H) basis sets employing the CCSD(T) method with excluded 1s$^2$2s$^2$2p$^6$ electrons of Lu. As one can see from \tref{TResult1}, the ratio
	$A_{||}/A_{\perp}$ is almost independent of the basis set. Hence, the values of $A_{\perp}/g_{\rm Lu}$ in \tref{TResult2} were obtained by the scaling of the corresponding values of $A_{||}/g_{\rm Lu}$ with the factor $A_{\perp}/A_{||} \approx 0.966$.
	
	Contribution of the Gaunt interaction has been calculated within the AE3Z basis set for Lu and the CV2Z \cite{Dyall:2016} basis set for O and H atoms. For this calculation we have used the code, developed in Ref. \onlinecite{Maison:2019} for atoms and generalized on the molecular case in the present paper. For the system under consideration the Gaunt correction, obtained within the CCSD(T) approach, is $+0.0135 \ \cdot 10^{33}\  \textrm{Hz}/(e\cdot \textrm{cm}^2)$  for $W_M$ and $-25$~MHz for $A_{||}/g_{\rm Lu}$.
	At the Dirac-Fock-Gaunt level it is $+0.0126\  \cdot  10^{33} \textrm{Hz}/(e\cdot \textrm{cm}^2)$ and  $-22$ MHz, respectively.
	
	It can be seen from Table~\ref{TResult2} that the noniteractive triple cluster amplitudes contribute less than 2\% to the $W_M$ value and about 1\% to the hyperfine structure constant value. Both Gaunt and basis set corrections (see also Table~\ref{TResult1}) contribute no more than 1.5\%. Note, that we do not consider the finite nuclear magnetization distribution correction contribution to the HFS constants which can achieve several percents~\cite{Skripnikov:2020e,Prosnyak:2020}. Basing on this analysis one can conclude that the uncertainty of the calculated $W_M$ value can be estimated to be lower than 5\%.

	\begin{table}[!h]
		\caption{
			Calculated $W_M$, $A_{||}/g_{\rm Lu}$ and $A_{\perp}/g_{\rm Lu}$ constants for the ground electronic state of the LuOH$^+$ molecular cation.}
		\label{TResult2}
		\begin{tabular}{llll}
			\hline
			\hline
			Contribution & $W_M$,  10$^{33} \frac{\textrm{Hz}}{e\cdot \textrm{cm}^2}$ & $A_{||}/g_{\rm Lu}$, MHz & $A_{\perp}/g_{\rm Lu}$, MHz\\
			\hline
			Dirac-Fock & $-$1.120 & 10409 & 10055\\  
			79e-CCSD & $-$1.294 &  12897 & 12456 \\
			\\
			79e-CCSD(T) &  $-$1.268 & 12785  & 12349 \\
			Gaunt correction &  +0.014 & $-$25 & $-$25 \\ 
			Basis set \\
			correction & +0.003 & $+$3  & $+$3\\
			\hline
			Final result & $\boldsymbol{-1.251}$& $\boldsymbol{12763}$ & $\boldsymbol{12327}$\\
			\hline
			\hline
		\end{tabular}
	\end{table}
	
	The final $W_M$ value is about 16\% higher than $W_M$(YbOH) \cite{Maison:2019b,denis2020enhanced}. 
	Note that the difference in constants of fundamental symmetry violating interactions in a neutral molecule and isoelectronic cation can be even larger~\cite{Skripnikov:15b,Skripnikov:2020c}.
	
	The resulting energy shift caused by the interaction (\ref{HmqmOperator}) can be parameterized in the following way \cite{Skripnikov:14a}:
	\begin{equation}
	\delta E = C(J,F,\Omega) \cdot |W_M M|,
	\label{deltaE}    
	\end{equation}
	with the parameter $C(J,F,\Omega)$ dependent on the sublevel of hyperfine structure considered; $J$ and $F$ are the the rotational level number and the total angular momenta respectively. For the ground and the lowest excited hyperfine structure levels this parameter may be estimated as 0.1 \cite{Skripnikov:14a,Petrov:18b,Kurchavov:2020}. Thus, one can expect the energy shift caused by CP-odd constants $\bar{\theta}$ and $(\tilde{d}_u  - \tilde{d}_{d})$ in $^{175}$LuOH$^+$ to be respectively
	\begin{eqnarray}
	\delta E(\bar{\theta}) \approx 20 \cdot 10^{10} \bar{\theta} \ \mu\textrm{Hz}, \\
	\delta E(\tilde{d}_u  - \tilde{d}_{d}) \approx 9 \cdot 10^{27} \ 
	\frac{\tilde{d}_u-\tilde{d}_{d}}{\textrm{cm}} \ \mu \textrm{Hz}.
	\end{eqnarray}
	
	Substitution of current constraints for $\bar{\theta}$ and $(\tilde{d}_u  - \tilde{d}_{d})$, taken from Ref. \onlinecite{Hg} ($|{\tilde \theta}| < 2.4 \cdot 10^{-10}$, $|{\tilde d_u}{-}{\tilde d_d}|~<~6~\cdot~10^{-27} $~cm), leads to the upper limits 
	$|\delta E(\bar{\theta})| \leq 48~\mu \textrm{Hz}$ and $|\delta E(\tilde{d}_u  - \tilde{d}_{d})| \leq 54~\mu \textrm{Hz}$. 
	\tref{Results for diatomics} compares the sensitivities of the T,P-violating energy shift to CP-violating parameters in LuOH$^+$ and molecules that are currently considered for the experimental search of T,P-violating effects. It is assumed that the corresponding parameter ($\bar{\theta}$ or $\tilde{d}_u  - \tilde{d}_{d}$) is the only source of the CP-violation. For all systems we set $C(J,F,\Omega)\approx 0.1$ in Eq.~(\ref{deltaE})
	for estimation of the energy shift $\delta E$. For comparison, \tref{Results for diatomics} provides also experimentally achieved constraints on the  T,P-violating energy shifts. Note, that these experiments have been performed with spinless isotopes of heavy nuclei, where the MQM is zero. The MQM and $W_M$ constants have also been estimated for a number of other molecules~\cite{Fleig:16a,Skripnikov:15c,Skripnikov:15cErratum,Talukdar:2019,talukdar2020role}. Note that the overall sensitivity to the mentioned CP-violating parameters for these molecules is smaller than for the $^{173}$YbOH molecule.

	As one can see, the expected energy shift for LuOH$^+$ is of the same order of magnitude as the current sensitivity achieved in measurements of the energy shift produced by the electron electric dipole moment in the ThO molecule \cite{ACME:18} and an about order of magnitude below the sensitivity of the HfF$^+$ experiment~\cite{Cornell:2017}. 
	The sensitivity of this experiment was in part limited by the achievable coherence time of 700 ms, which was due to collisions between ions. This decoherence mechanism can be suppressed by decreasing ion temperature, either by direct laser cooling, or sympathetic cooling. Coherence times as long as 10 minutes have been demonstrated in a single sympathetically cooled ion~\cite{Wang2017}. As an example, a sensitivity of 48 $\mu$Hz could be achieved in an experiment using a single LuOH$^+$ cation, sympathetically cooled by a laser cooled Lu$^+$ ion, with a 30 second spin precession time, and 100 hours of averaging. Taking also into account the large coherence time one may expect that LuOH$^+$ promises to give new restrictions for mentioned CP-odd fundamental parameters.
	
	%dm_resp
	\begin{table}[]
		%\centering
		\caption{
			%ls   
			The values of $\delta E/\bar{\theta}$ and $ \delta E/(\tilde{d}_u - \tilde{d}_d)$ for several molecules and cations with heavy-atom nuclei ($\delta E\approx 0.1 W_M M$). For comparison, current experimental constraints on the T,P-violating energy shift are given where available (for spinless isotopes).
		}
		\begin{tabular}{lllll}
			\hline
			Molecule & $|W_M|$, & $|\delta E / \bar{\theta}|$, &
			$ |\delta E/(\tilde{d}_u - \tilde{d}_d)|$,  & $|\delta E|$,
			\\
			& $10^{33} \frac{\textrm{Hz}}{e\cdot \textrm{cm}^2}$  & $10^{10}\mu \textrm{Hz}^a$ &  
			$10^{27}\mu \textrm{Hz/cm}^a$  &$\mu$Hz \\
			\hline
			\hline
			$^{174}$YbF ($I$=0)& 
			\multirow{2}{*}{1.2$^b$} & & & 3500 \cite{Hudson:11a} \\
			$^{173}$YbF ($I$=5/2)&  & 10 & 5  \\
			\hline
			$^{180}$HfF$^+$ ($I$=0)& \multirow{3}{*}{0.5$^c$}& & &700\cite{Cornell:2017}\\ 
			$^{177}$HfF$^+$ ($I$=7/2)&  & 4 & 2 \\
			$^{179}$HfF$^+$ ($I$=9/2)&  & 4 & 2 \\
			\hline
			$^{232}$ThO ($I$=0) &\multirow{2}{*}{1.1$^{d}$}  &  &  & 220\cite{ACME:18}\\
			$^{229}$ThO ($I$=5/2)&  & 5 & 2 \\
			\hline
			$^{137}$BaF ($I$=3/2)& 0.4$^e$ & 0.1 &0.06  & \\
			\hline
			$^{181}$TaO$^+$ ($I$=7/2)& 0.45$^f$ & 3 & 2 & \\
			\hline
			$^{232}$ThF$^+$ ($I$=0) & \multirow{2}{*}{0.6$^g$}& & & \\
			$^{229}$ThF$^+$ ($I$=5/2) & & 3 & 1 & \\
			\hline
			$^{173}$YbOH ($I$=5/2)& 1.1$^h$ & 10 & 6 \\
			\hline
			$^{175}$LuOH$^{+}$ ($I$=7/2)& \multirow{2}{*}{1.3$^i$}& 20 & 9 \\
			$^{176}$LuOH$^{+}$ ($I$=7)  & & 28 & 12 \\
			\hline
			\hline
		\end{tabular}
		\label{Results for diatomics}
		\begin{flushleft}
			$^a$Refs. \onlinecite{lackenby2018time,FDK14}
			\\
			$^b$Refs. \onlinecite{Titov:96b,Quiney:98,denis2020enhanced}
			\\
			$^c$Ref. \onlinecite{Skripnikov:17b}
			\\
			$^d$Refs. \onlinecite{Skripnikov:14a,Skripnikov:14aa}
			\\
			$^e$Refs. \onlinecite{KL95,denis2020enhanced}
			\\
			$^f$Refs. \onlinecite{Fleig:2017}
			\\
			$^g$Ref. \onlinecite{Skripnikov:15b,Skripnikov:17b}
			\\
			$^h$Refs.\onlinecite{Maison:2019b,denis2020enhanced}
			\\
			$^i$This work.
		\end{flushleft}
	\end{table}

	\begin{acknowledgments} 
		We are  grateful to Igor Samsonov and Alexander Oleynichenko for useful discussions. 
		Electronic structure calculations have been carried out using computing resources of the federal
		collective usage center Complex for Simulation and Data Processing for
		Mega-science Facilities at NRC “Kurchatov Institute”, http://ckp.nrcki.ru/.
		
		$~~~$Molecular coupled cluster electronic structure calculations have been supported by the Russian Science Foundation Grant No. 18-12-00227. Calculations of the $W_M$, $A_{||}$ and $A_{\perp}$ matrix elements were supported by the foundation for the advancement of theoretical physics and mathematics ``BASIS'' grant according to the research projects No. 18-1-3-55-1 and No. 20-1-5-76-1. Calculation of the Gaunt interaction matrix elements has been supported by RFBR grant No. 20-32-70177. The calculations of the nuclear structure were supported by the Australian Research Council Grants No. DP150101405, DP200100150 and New Zealand Institute for Advanced Study. M. G. acknowledges support from the Swiss National Science Foundation Grant No. BSCGI0 157834.
	\end{acknowledgments} 
	
	\section*{data availability}
	The data that supports the findings of this study are available within the article.

	\bibliographystyle{apsrev}
	%\bibliographystyle{aipsamp}
	%\bibliography{JournAbbr,SkripnikovLib,QCPNPI,TitovLib,Maison,VictorFlambaum,Grau}

\end{document}